\documentclass[11pt]{article}

\usepackage[hyphens]{url}
\usepackage[margin=1in]{geometry}
\usepackage[toc]{appendix}
\usepackage{graphicx}
\RequirePackage[OT1]{fontenc}
\usepackage{natbib}
\usepackage{adjustbox}
\usepackage{amsmath,amssymb,amsthm,bm,latexsym}
\usepackage{graphicx,psfrag,epsf}
\usepackage{epigraph,xcolor}
\usepackage{enumerate}

\usepackage{booktabs}
\usepackage{multirow}
\usepackage{hyperref}
\usepackage{float}

\usepackage{algpseudocode}
\usepackage[linesnumbered,ruled]{algorithm2e}
\hypersetup{colorlinks=true,citecolor=blue,urlcolor=purple,
	pdfpagemode=FullScreen,linktocpage=true}
\usepackage{cleveref}

\allowdisplaybreaks

\numberwithin{equation}{section}

\newcommand{\rd}{{\rm d}}

\begin{document}

\title{Modeling Motion Dynamics in Psychotherapy: a Dynamical Systems Approach}

\author{Itai Dattner\\Department of Statistics, University of Haifa\\idattner@stat.haifa.ac.il}

\maketitle

\begin{abstract}
This study introduces a novel mechanistic modeling and statistical framework for analyzing motion energy dynamics within psychotherapy sessions. We transform raw motion energy data into an interpretable narrative of therapist-patient interactions, thereby revealing unique insights into the nature of these dynamics. Our methodology is established through three detailed case studies, each shedding light on the complexities of dyadic interactions. A key component of our approach is an analysis spanning four years of one therapist's sessions, allowing us to distinguish between trait-like and state-like dynamics. This research represents a significant advancement in the quantitative understanding of motion dynamics in psychotherapy, with the potential to substantially influence both future research and therapeutic practice.

\end{abstract}

\section{Introduction}
Psychotherapy is a vital tool in the management and treatment of various mental health disorders, and its effectiveness is dependent on a multitude of factors \citep{lambert2004efficacy}. Among these, the quality of the therapeutic alliance between the patient and the therapist is a crucial determinant of the treatment outcome \citep{horvath2011alliance}. Research in this domain has highlighted the significance of non-verbal cues, such as body language and facial expressions, as essential components of the therapeutic relationship \citep{ramseyer2011nonverbal}. The phenomenon of non-verbal synchrony, where the patient and therapist unconsciously mirror each other's movements, is of particular interest, as it is associated with positive therapeutic outcomes (\cite{ramseyer2011nonverbal}, \cite{koole2016synchrony}). Traditionally, such non-verbal information has been studied using observational methods, relying on human coders to evaluate the degree of movement similarity between the dyadic partners. These approaches are time-consuming, subjective, and prone to human error. The advent of motion capture technology and advanced computational techniques has paved the way for more reliable and objective assessments of non-verbal information. Motion energy analysis (MEA) is one such promising approach that quantifies the spatial and temporal patterns of movement in a dyad. The MEA software generates data that capture the movement patterns of participants in a dyadic interaction, such as those occurring during psychotherapy sessions. These data are obtained by processing video recordings of the sessions, where the software extracts and quantifies the motion energy present in the temporal sequences. The resulting output consists of time series data that represent the movement intensity for each individual in the dyad. Specifically, 
frame-differencing algorithms quantify movement dynamics by measuring differences between consecutive frames in a sequence . These algorithms compare each frame to its predecessor and extract the differences based on the number and magnitude of pixel changes. While frame-differencing methods effectively quantify the degree of change over time, they do not capture the direction or form of movement, as they solely focus on the extent of change between frames. 

\begin{figure}[!ht]
\centering
\includegraphics[width=0.8\textwidth]{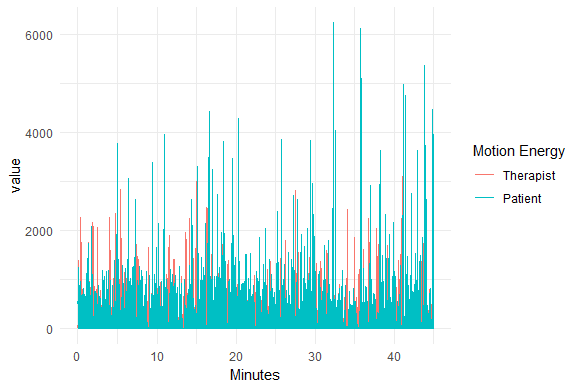}
\caption{\label{fig:ME1} A 45-minute segment of therapist and patient motion energy data obtained from the MEA software. The segment, spanning from the 10th to the 55th minute of each session, excludes initial logistical discussions and focuses on the core therapeutic interaction.}
\end{figure}

In this study we use data obtained from MEA, which are publicly accessible as detailed in \cite{Ramseyer_2023}. These data, thoroughly studied by \cite{ramseyer2020motion}, allows us to implement and evaluate our innovative methods and inference framework. The data showcase therapist and patient movement patterns during sessions. Figure \ref{fig:ME1} depicts a 45-minute segment of therapist and patient motion energy data, derived from the MEA software. This segment doesn't include the initial ten minutes of each session, which typically involve logistical discussions such as setting up video recording and discussing financial details. The substantive portion of the session usually begins with a question about the patient's motivation for seeking the appointment. The decision to exclude the first ten minutes ensures that the organizational aspects of the session are not part of the analysis, focusing instead on the psychotherapeutic interaction. The selected segment for analysis, therefore, spans from the 10th to the 55th minute of the session, providing a consistent timeframe for all sessions analyzed in the sequel. Motion energy time series such as the one displayed in the figure, serves as the basis for our exploration of therapist and patient motion dynamics.

Although MEA has been applied in various contexts, including psychotherapy research , its full potential remains untapped, as current models do not fully capture the intricate motion dynamics within a dyad. There is a pressing need for novel mathematical models that can elucidate the complex interplay of motion dynamics and offer a more comprehensive understanding of the underlying mechanism. 
We use the notation $x_0(t)=(x_{01}(t),x_{02}(t))^\top$ for the continuous movement velocities of the therapist and patient, respectively; $\top$ stand for the transpose of a vector. In what follows we use the terms 'motion energy' and 'velocity' interchangeably. Without loss of generality, we use $t\in[0,1]$ to represent a full session of 45 minutes. Denote the vector of derivatives of $x_0(t)$ w.r.t. $t$ by
\begin{equation}
\label{eq:f}
f_0(t)=(x^\prime_{01}(t),x^\prime_{02}(t))^\top, \quad t\in[0,1]. 
\end{equation}

The scientific question posed in this work is essentially a question of finding an adequate parametric description for $f_0(\cdot)$ defined in Equation (\ref{eq:f}), one that expresses $f_0(t)$ in terms of $x_1(t), x_2(t)$ that describes the process mechanistically, in the sense that the current rate-of-change depends on the current state. In undertaking this study, we are cognizant that human motion within a psychotherapy session is an inherently complex phenomenon, intricately entwined with a multitude of psychological, emotional, and contextual factors. It is, therefore, important to acknowledge that any model we propose is an approximation of reality and, in this sense, is inherently misspecified. However, the goal of our work is not to propose a perfect model that captures all facets of motion dynamics, but rather to develop a useful model that affords us insights into the underlying mechanisms governing these dynamics. 

The structure of the paper is as follows. Section 2 introduces the  ordinary differential equation model and the corresponding mechanistic implications. Section 3 establishes the statistical framework, while an empirical analysis of three specific dyadic interactions  is the topic of Section 4. 
A key result is developed in Section 5 where we distinguish between trait and state characteristics in motion dynamics. Finally, Section 6 discusses findings and future research directions. This work aims to integrate mathematical modeling, statistics, and psychotherapy research to better understand and quantify  motion dynamics in psychotherapy dyadic interactions. 


\section{Mechanistic Modeling of Motion Dynamics}

The exploration of motion dynamics in psychotherapy interactions can be significantly enriched by considering not just velocity but also acceleration—the rate of change in velocity. The analysis of acceleration, a derivative of velocity, affords a detailed understanding of the temporal dynamics of motion, potentially revealing subtle alterations in movement patterns that are otherwise missed when focusing solely on velocity. For example, in the engineering domain, acceleration is used as an input to better control  complex dynamics, see, e.g., \cite{nise2020control}. Applying this understanding to psychotherapy, therapists' awareness of acceleration changes in their own and their patients' movements could serve as a novel 'control mechanism.' This could enable therapists to better guide the therapeutic alliance by mirroring or complementing their patients' non-verbal cues, thereby enhancing rapport and mutual understanding. Thus, we leverage dynamical systems theory to develop novel methodologies for assessing motion dynamics within psychotherapy dyadic interactions. We consider a parametric model given by a coupled system of ordinary differential equations (ODEs). The linear ODEs system is given by 

\begin{equation}
\label{eq:ode}
\begin{cases}
x_1^{\prime}(t)=\alpha x_1(t)+\beta x_2(t),\\
x_2^{\prime}(t)=\gamma x_1(t)+\delta x_2(t),
\end{cases}
\end{equation}

which encodes the dynamics of a psychotherapy session. Here, the states $x_1(t)$ and $x_2(t)$ represent the therapist's and patient's motion energy levels, respectively. Their temporal derivatives, $x_1^{\prime}(t)$ and $x_2^{\prime}(t)$, capture the movements acceleration or deceleration. The ODEs are characterized by their ability to model interactions between multiple entities or processes, making them particularly suitable for capturing the complex interplay between therapist and patient during a psychotherapy session, see e.g., \cite{tschacher2019process}.

The above coupled linear system of ODEs 
has been thoroughly studied, and its  qualitative properties, which include stability, periodicity, and sensitivity to initial conditions, among others, are well understood. In particular, the analytic solution to the system of ODEs \eqref{eq:ode} can be obtained via matrix exponentiation: 
\[
\begin{aligned}
\begin{bmatrix} x_1(t) \\ x_2(t) \end{bmatrix} = e^{At}\begin{bmatrix} x_1(0) \\ x_2(0) \end{bmatrix},
\end{aligned}
\]

where \(A\) is the matrix of coefficients:

\[
\begin{aligned}
A = \begin{bmatrix} \alpha & \beta \\ \gamma & \delta \end{bmatrix}.
\end{aligned}
\]
Here \(e^{At}\) is the matrix exponential, which can be computed using a series expansion or via eigendecomposition. The specific form of the solution depends on the eigenvalues and eigenvectors of \(A\). The eigenvalues \(\lambda_1\) and \(\lambda_2\) of \(A\) are the solutions to the characteristic equation, which is given by $\text{det}(A - \lambda I) = \lambda^2 - (\alpha + \delta) \lambda + (\alpha \delta - \beta \gamma) = 0$; here $I$ is the identity matrix. For instance, in case of real and distinct eigenvalues (\(\lambda_1 \neq \lambda_2\)) the general solution will be of the form
   \[
   \begin{aligned}
   \begin{bmatrix} x_1(t) \\ x_2(t) \end{bmatrix} = c_1 e^{\lambda_1 t} \mathbf{v}_1 + c_2 e^{\lambda_2 t} \mathbf{v}_2
   \end{aligned},
   \]   
where \(c_1\) and \(c_2\) are constants determined by the initial conditions, and \(\mathbf{v}_1\) and \(\mathbf{v}_2\) are the eigenvectors corresponding to \(\lambda_1\) and \(\lambda_2\), respectively. On the other hand, if the eigenvalues are complex, they will come in complex conjugate pairs, \(\lambda_{1,2} = a \pm bi\), and the general solution will be of the form
   \[
   \begin{aligned}
   \begin{bmatrix} x_1(t) \\ x_2(t) \end{bmatrix} = e^{at} (c_1 \cos(bt) + c_2 \sin(bt)) \mathbf{v}
   \end{aligned}.
   \]   
Here \(c_1\) and \(c_2\) are constants determined by the initial conditions, and \(\mathbf{v}\) is the eigenvector corresponding to \(\lambda_{1,2}\). In each case, the specific form of the solution and its behavior over time will depend on the values of the parameters \(\alpha\), \(\beta\), \(\gamma\), and \(\delta\). The parameter $\alpha$ is a self-damping/reinforcing term for the therapist, indicating the rate at which the therapist's motion energy tends to stabilize/de-stabilize, respectively. The coefficient $\alpha$ multiplies the term $x_1(t)$ in the equation for $x_1^{\prime}(t)$. As such, $\alpha x_1(t)$ describes the component of the therapist's motion energy change that depends solely on the therapist's current motion energy level. A negative/positive $\alpha$ suggests a damping/reinforcing effect, with the therapist's motion energy decelerating/accelerating as it increases. For instance, when $\alpha$ is negative this might represent a more reserved therapeutic style or a more structured therapeutic approach. The magnitude of $\alpha$ modulates this effect: a larger (negative) magnitude implies a quicker return to a stable state. The parameter $\delta$ serves a similar role for the patient, being the self-damping/reinforcing factor that governs how quickly the patient's motion energy tends to stabilize/de-stabilize.

On the other hand, the parameters $\beta$ and $\gamma$ reflect cross-influences between the therapist and patient. A positive $\beta$ implies that an increase in the patient's motion energy tends to raise the therapist's motion energy, indicating a synchronous dynamic. Conversely, a negative $\beta$ suggests a counterbalancing dynamic, where the patient's increased motion energy tends to decelerate the therapist's motion energy. The parameter $\gamma$ mirrors this dynamic, but with the roles of the therapist and patient reversed.

In our analysis, we also consider transformations of the model parameters into a set of ratios that represent the relative contribution of each factor within the psychotherapy session. This transformation allows for a more interpretable insight into the dyadic dynamics. The transformed parameters are:

\begin{enumerate}
    \item \textbf{Therapist self-damping/reinforcing ratio ($th_{self}$):} This is defined as the absolute value of $\alpha$ divided by the sum of the absolute values of all parameters. Mathematically, this is expressed as $th_{self}=|\alpha|/(|\alpha|+|\beta|+|\gamma|+|\delta|)$.
    \item \textbf{Therapist interaction ratio ($th_{int}$):} Similar to the damping/reinforcing ratio, the therapist interaction ratio is calculated as the absolute value of $\beta$ divided by the sum of the absolute values of all parameters: $th_{int}=|\beta|/(|\alpha|+|\beta|+|\gamma|+|\delta|)$.    
    \item \textbf{Patient interaction ratio ($pa_{int}$):} The patient interaction ratio is the absolute value of $\gamma$ divided by the sum of the absolute values of all parameters: $pa_{int}=|\gamma|/(|\alpha|+|\beta|+|\gamma|+|\delta|)$.
    \item \textbf{Patient self-damping/reinforcing ratio ($pa_{self}$):} This ratio is calculated  for the $\delta$ parameter: $pa_{self}=|\delta|/(|\alpha|+|\beta|+|\gamma|+|\delta|)$.
\end{enumerate}

Incorporating these parameters and their respective ratio definitions, our model provides a nuanced understanding of psychotherapy dyadic interactions. It accounts not just for the motion energy levels but also for their rates of change - acceleration and deceleration. The four ration parameters allow us to quantify the specific contribution of each participant's motion dynamics to the overall session. With these parameters and ratios, we can better understand and interpret the motion dynamics and non-verbal synchrony observed in therapist-patient interactions.

As mentioned above, model misspecification is an important consideration in statistical analysis. In our study, we use a system of ordinary differential equations to model the dynamics of motion energy within psychotherapy sessions. If the solution of this ODE system precisely represents the true motion velocities, then the model is well specified. However, if the solution only approximates the true velocities, then the model is, in fact, misspecified. Under such circumstances, the parameters of the ODE system, say $\theta:=(\alpha,\beta,\gamma,\delta)^\top$ and initial values $\xi:=(x_1(0),x_2(0))^\top$, do not necessarily correspond to the "true" parameters but rather those giving rise to solutions that are closest to the true velocities in the following sense:
\begin{equation}\nonumber
(\theta,\xi):=\arg\min_{\theta\in\Theta,\xi\in\Xi}\int_0^1 \left|\left|
x(\theta,\xi;t)-x_0(t)
\right|\right|^2\,\rd t,
\end{equation}
where $\parallel \cdot \parallel$ denotes the Euclidean norm.
This interpretation is based on the work of \cite{white1982maximum} on model misspecification. In the presence of model misspecification, the parameters of the ODEs are still interpretable, but their interpretations may differ from the case where the model is correctly specified. This caveat should be borne in mind when interpreting the estimated parameters and the model's insights. However, the potential for model misspecification does not undermine the utility of our approach. It merely provides a reminder of the need for careful interpretation of the results and the context-dependent nature of the parameter estimates.

\section{Statistical Inference Framework}

The observed time series of motion energy consist of positive values over a session as displayed in Figure~\ref{fig:ME1}. The data for our study were sourced from video-recorded intake interviews at a psychotherapy clinic in Bern, Switzerland \citep{ramseyer2020motion}. Each interview lasted between 60 to 90 minutes, with a focus on understanding the patient's reasons for seeking therapy and their personal life history. To ensure consistency in our analysis, we omitted the initial ten minutes of each session, typically spent on logistics, and confined our analysis to a 45-minute segment starting from the 10th minute.

\subsection{Statistical Model}

We split a session into ten equidistant segments and summarize each segment by the mean motion energy values denoted by $Y_{j}(t_i)$, $i=1,...,n$, $j=1,2$ (in our case $n=10$). This is a noisy version of the underlying mean of the true motion energy process denoted above by $x_{0j}(t_i)$. After standartization of these values we consider the statistical model 
\begin{equation}\nonumber
Y_{j}(t_i)=x_{0j}(t_i)+\epsilon_{ij}, \quad i=1,\ldots,n \quad j=1,2,
\end{equation}
where $Y_j(t_i)$ is a scalar random variable,   $t_1,\ldots,t_n$ are deterministic distinct design points; and the unobserved random variables $\epsilon_{ij}$ are independent measurement errors having zero expectation and finite variance. There is some abuse of notation that simplifies presentation where we use $x_{0j}$ in the observation model which is now considered to be the standardized underlying process.

The choice to segment psychotherapy sessions into ten equidistant periods for model fitting was informed by existing practices in the field. For instance, in the psychotherapy research literature, it is common for human coders to divide sessions into five-minute segments, assigning various clinical labels to each of these segments based on the observed dynamics (see, e.g., \cite{deres2021nonverbal}). Here we work with 45 minutes so each segment is of 4.5 minutes. This approach allows for a more granular understanding of the therapeutic process, capturing the evolving nature of therapist-patient interaction and potential shifts in clinical dynamics throughout the session. Our type of motion energy data are inherently noisy, and by taking averages over short intervals, we ensure that our estimates better represent the underlying process. This technique reduces the impact of momentary fluctuations and enhances the reliability of our parameter estimates, despite the noise in the raw data. We have also studied the finite sample properties of our method. The Monte Carlo simulation results, as seen in Table 1 and Table 2, suggest that our choice of ten segments is not only practical but also statistically reliable, even in the face of measurement error.

By aligning our model fitting process with this established method of data organization, we position our modeling approach to readily incorporate these clinically-relevant labels whenever they become available. This ensures that our model, based on motion energy dynamics, and the clinical labels share the same temporal scale, facilitating a more integrated and clinically nuanced analysis, providing a potentially valuable tool for understanding and interpreting therapeutic processes.


\subsection{Parameter Estimation}

The next crucial step in our analysis is parameter estimation for our system of ordinary differential equations. This task forms the heart of the mechanistic modeling process, as it enables us to quantitatively characterize the dynamics of therapist-patient interactions. Recent advancements in the field have provided a suite of techniques for ODE parameter estimation, as comprehensively reviewed in \cite{dattner2021differential}. 

Let $\theta:=(\alpha,\beta,\gamma,\delta)^\top$ and note that the ODEs are equivalent to the integral equations
\begin{equation}\label{eq:int_model}
x(t)=\xi + \int_0^t g(x(s))\, \rd s\,\theta,\ t\in[0,1],
\end{equation}
where $\xi=(\xi_1,\xi_2)^\top$ stands for the initial values of the system $x(0)=(x_1(0),x_2(0))^\top$, and the matrix $g$ is given by
\[
\begin{aligned}
g(x) = \begin{bmatrix} x_1 & x_2 & 0 & 0\\ 0 & 0 &x_1 & x_2 \end{bmatrix}.
\end{aligned}
\]

Let
\begin{eqnarray}\nonumber
G(t) &=& \int_0^t g(x(s))\,\rd s,\,\quad \ t \in
[0,1],\nonumber \\\nonumber
A &=& \int_0^1 G(t)\rd t, \quad B= \int_0^1 G^T(t) G(t)\rd t.
\end{eqnarray}
\cite{dattner2015optimal} show that 
if $B$ is nonsingular then
$I-AB^{-1}A^T$ is and
\begin{eqnarray}\nonumber
\xi &=& \left(I - A B^{-1}A^T\right)^{-1} \int_0^1 \left(I - A
B^{-1} G^T(t)\right) x(t)\, \rd t, \label{xi} \\\nonumber
\theta &=& B^{-1} \int_0^1 G^T(t) \left( x(t) -\xi \right) \rd t
\label{theta}
\end{eqnarray}
hold; here $I$ denotes the $2 \times 2$ identity matrix. Moreover, they show that if $x(t)$ determines $\theta,$ then $B$
is nonsingular. This provides necessary and sufficient
conditions for identifiability of $\theta$. 

Note that the system of ODEs representing the motion dynamics within a psychotherapy session is linear in the parameters. This property is instrumental in the estimation process. While an analytic solution exists for the ODEs, as described above, we opt for minimizing the distance between observations and the ODEs' solution for parameter estimation. This leads to a nonlinear least squares problem. However, this can be circumvented by adopting the direct integral approach proposed by \cite{dattner2015optimal}; see also \cite{dattner2020separable}. The direct integral approach provides a more accurate and computationally efficient mechanism for estimating parameters of ODEs linear in (function) of the parameters. The aforementioned works have demonstrated the robustness and efficiency of this approach, making it a valuable tool for our analysis.

In order to estimate the parameter $\theta$ the observations are first smoothed, which
results in an estimator $\hat{x}_n(\cdot)$ for the solution
$x(\cdot;\theta,\xi)$ of the system, and by differentiation in the
estimator $\hat{x}_n^{\prime}(\cdot)$ for
$x^\prime(\cdot;\theta,\xi).$ Then in view of Equation (\ref{eq:int_model}) we estimate the parameters $\theta$ and $\xi$ by minimizing
\begin{equation}\label{criterion}
\int_0^1 \left|\left|
\hat{x}_n(t)-\zeta-\int_0^tg(\hat{x}_n(s))\,\rd s\,\eta
\right|\right|^2\,\rd t.
\end{equation}
Denote
\begin{eqnarray}\nonumber
\hat{G}_n(t) &=& \int_0^t g(\hat{x}_n(s))\,\rd s\,,\quad t \in
[0,1],\nonumber \\\nonumber
\hat{A}_n &=& \int_0^1 \hat{G}_n(t)\,\rd t, \\
\hat{B}_n &=& \int_0^1 \hat{G}_n^\top (t) \hat{G}_n(t)\,\rd t. \nonumber
\end{eqnarray}
Minimizing the criterion function (\ref{criterion}) with respect
to $\zeta$ and $\eta$ results in the direct estimators
\begin{eqnarray}
\hat{\xi}_n &=& \left(I - \hat{A}_n \hat{B}_n^{-1}
\hat{A}_n^\top\right)^{-1} \int_0^1 \left(I - \hat{A}_n
\hat{B}_n^{-1}
\hat{G}_n^\top (t)\right) \hat{x}_n(t)\,\rd t, \label{xihat} \\
\hat{\theta}_n &=& \hat{B}_n^{-1} \int_0^1 \hat{G}_n^\top (t) \left(
\hat{x}_n(t) -\hat{\xi}_n \right) \rd t. \label{thetahat}
\end{eqnarray}
 \cite{dattner2015optimal} present conditions that
guarantee $\sqrt n$-consistency of the estimators
$\hat{\xi}_n$ and $\hat{\theta}_n.$

\subsection{Finite Sample Properties}
Following the introduction of our estimation method, we now turn our attention to assessing the finite sample properties of our estimators. Given the structure of our data, with each psychotherapy session being divided into ten equidistant segments, it becomes crucial to understand how our estimators behave under this finite sample scenario. Thus, Monte Carlo simulations are integral to our study, offering invaluable insights into the distributional properties of our estimators under controlled conditions. This small numerical study aims to ensure our findings' reliability and validity, laying a solid groundwork for future psychotherapy research applications of this methodology. However, it's a targeted inquiry driven by our modeling approach's practical considerations, not a comprehensive exploration of the estimator’s properties. More comprehensive Monte Carlo studies and a thorough theoretical investigation of the estimators defined in Equations \eqref{xihat}-\eqref{thetahat} have been conducted in the foundational work by \cite{dattner2015optimal}. For implementation we employ the `simode` package in R developed by \cite{yaari2019simode}. The `simode` package is particularly well-suited for our needs, as it is designed to handle ODE models that are linear in their parameters, which aligns perfectly with the structure of our model. This package utilizes state-of-the-art techniques to reliably estimate model parameters, even in complex settings. Using `simode`, we can extract meaningful parameter values from our preprocessed motion energy data, thereby providing a quantitative foundation for understanding the dyadic dynamics within psychotherapy sessions. 

Guided by the real data analysis presented in the sequel we study the case of a sample size of $n=10$. We set the parameters to $\alpha=-0.5,\beta= -1.5,\gamma= 1,\delta= 0.3$. The initial conditions for the ODEs were set at $0.5, -0.5$ for the therapist and patient, respectively. We added Gaussian noise to the ODE solutions to simulate measurement errors at two levels, 20\% and 50\% of the mean value of the solution. The time interval for the simulation ranged from 1 to 10, and the simulations were repeated 100 times.

In the case of a 20\% measurement error, as shown in Table~\ref{tab:table1}, the estimated means for the therapist and patient parameters come remarkably close to the truth, providing excellent results. The variance is also suitably managed, ensuring reliable estimations. The estimation of the damping/reinforcing and interaction ratios is particularly accurate, demonstrating the robustness of our method.
\begin{table}[h!]
\centering
\caption{Results of Monte Carlo simulations with a sample size of 10 and measurement error of 20\%, showing estimated mean (standard deviation) of the parameters and the ratios.}
\label{tab:table1}
\begin{tabular}{ccc}
\toprule
Parameter & True Value & Estimated Mean (Std) \\
\midrule
$ \alpha $ & -0.5 & -0.56 (0.01) \\
$ \beta $ & -1.5 & -1.47 (0.01) \\
$ \gamma $ & 1 & 1.00 (0.01) \\
$ \delta $ & 0.3 & 0.34 (0.01) \\
$ \text{th\_self} $ & 0.15 & 0.17 (0.00) \\
$ \text{th\_int} $ & 0.45 & 0.44 (0.01) \\
$ \text{pa\_int} $ & 0.30 & 0.30 (0.00) \\
$ \text{pa\_self} $ & 0.09 & 0.10 (0.00) \\
\bottomrule
\end{tabular}
\end{table}

When the measurement error is increased to 50\%, the results, although somewhat affected, remain promising. As Table~\ref{tab:table2} illustrates, even with a larger measurement error and a small sample size of 10, the estimated means for parameters are close to the truth. The variance, while present, is not prohibitive for reliable estimation. The estimation of the self-damping/reinforcing and interaction ratios remains reasonably accurate, further underlining the efficacy of our approach. These results illuminate the potential of our method to consistently estimate these pivotal ratios, providing a more interpretable and significant understanding of the dynamics in psychotherapy.

\begin{table}[h!]
\centering
\caption{Results of Monte Carlo simulations with a sample size of 10 and measurement error of 50\%, showing estimated mean (standard deviation) of the parameters and the ratios.}
\label{tab:table2}
\begin{tabular}{ccc}
\toprule
Parameter & True Value & Estimated Mean (Std) \\
\midrule
$ \alpha $ & -0.5 & -0.55 (0.03) \\
$ \beta $ & -1.5 & -1.47 (0.04) \\
$ \gamma $ & 1 & 1.00 (0.03) \\
$ \delta $ & 0.3 & 0.34 (0.03) \\
$ \text{th\_self} $ & 0.15 & 0.16 (0.01) \\
$ \text{th\_int} $ & 0.45 & 0.44 (0.01) \\
$ \text{pa\_int} $ & 0.30 & 0.30 (0.01) \\
$ \text{pa\_self} $ & 0.09 & 0.10 (0.01) \\
\bottomrule
\end{tabular}
\end{table}

\section{Empirical Analysis of Dyadic Interactions in Psychotherapy Sessions using a Mechanistic Model}

In this section we apply our proposed mechanistic model to real-world data obtained from psychotherapy sessions. This empirical analysis aims to elucidate the dynamic interactions between therapists and patients during these sessions, providing a novel quantitative perspective on the psychotherapeutic process. The motion energy data, derived from video recordings of the sessions, serves as a proxy for the movements velocities of the involved individuals. By fitting our model to these data, we seek to capture the inherent dyadic dynamics and provide a robust framework for exploring various hypotheses related to psychotherapy practice. Herein, we present the results of this empirical analysis, highlighting key findings and interpreting them in the context of psychotherapeutic interaction. 

In our analysis of motion energy data, we specifically focused on the regions of interest (ROI) corresponding to the heads of the therapist and the patient. This decision was based on the capabilities of the software used to capture and process the motion energy data, which allows for the definition of specific ROIs. By concentrating on the head regions, we aimed to capture a significant portion of the expressive behavior and nonverbal communication cues often crucial in psychotherapy sessions. This includes various head movements and postures which can convey agreement, attention, emotion, and other psychological states. Therefore, the data derived from these ROIs present a rich source of information for understanding the dynamic interaction between the therapist and patient.

\subsection{Data Preprocessing}
As a preliminary step in the data processing pipeline, each psychotherapy session is partitioned into ten equal segments from reasons detailed above. We then calculate the mean motion energy for each of these segments, which provides a coarse-grained representation of the activity levels throughout the session. To account for potential differences in baseline activity levels across different sessions or individuals, these mean motion energy values are standardized. This ensures that the values for each segment reflect deviations from the average activity level, rather than absolute measures of motion energy. Thus, through this preprocessing pipeline, we ensure that the dataset is adequately prepared for the application of the mechanistic model, allowing us to capture the essential dynamics of therapist-patient interactions within each session.

\subsection{Case Studies: In-Depth Analysis of Single Psychotherapy Sessions}

Psychotherapy sessions are complex, involving myriad subtle interactions and dynamics, and the intent of the following case studies is to illuminate these complexities in a way that brings our analytical approach to life. By focusing on individual sessions, we can draw out the nuances of our model and elucidate the meaningful interpretations of the parameters and ratios in the specific context of psychotherapy. This process provides the reader with a deeper understanding of our approach, laying the groundwork for the subsequent large-scale data analysis. The case studies, therefore, serve as a bridge, translating the abstract methodology into a practical framework for analysis, and setting the stage for the broader investigation that follows in the next section.

\begin{figure}[!ht]
\centering
\includegraphics[width=0.6\textwidth]{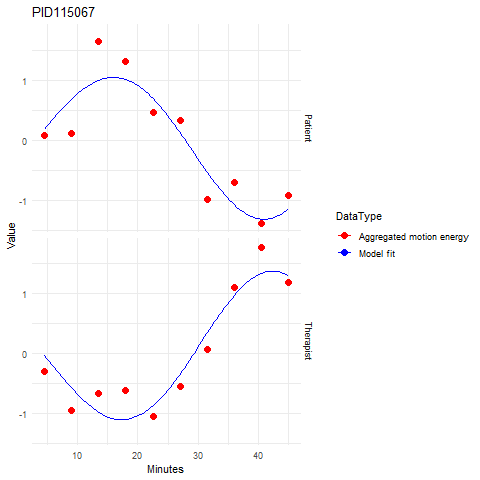}
\caption{\label{fig:fit1} Patient ID 115067: This figure depicts the aggregated motion energy data for both the therapist and patient, with a superimposed model fit. The model fit curve (shown in blue) demonstrates the performance of the model in capturing the patterns of the actual data, represented by red points. The plot provides a visual assessment of the model's accuracy in fitting motion energy trends.}
\end{figure}

In the first case study we analyze Patient ID 115067. The parameter estimates are $\hat{\alpha} = -0.433$, $\hat{\beta} = -0.478$, $\hat{\gamma} = 0.442$, and $\hat{\delta} = 0.452$. Here $\alpha$ and $\beta$ showing negative estimates, and the eigenvalues are complex conjugate which suggest an oscillatory pattern in the interaction dynamics, as can be seen in Figure~\ref{fig:fit1}. 
We also provide confidence intervals for the parameters using
profile likelihood generated by 'simode' package. 
The 95\% confidence intervals for the parameter estimates are provided in Table \ref{tab:confint1}. Notably, all the parameters have confidence intervals that exclude zero, indicating that they are statistically significant. The intervals are relatively narrow, reflecting a high degree of precision in these estimates. This suggests that the parameters are identifiable under the conditions of this analysis.
\begin{table}[ht]
\centering
\caption{Patient ID 115067: 95\% Confidence Intervals for Parameter Estimates.}
\begin{tabular}{lcccc}
\hline
Parameter & Estimate & Lower Bound & Upper Bound \\
\hline
$\alpha$ & -0.433 & -0.435 & -0.431 \\
$\beta$ & -0.478 & -0.528 & -0.476 \\
$\gamma$ & 0.442 & 0.440 & 0.488 \\
$\delta$ & 0.452 & 0.445 & 0.454 \\
\hline
\end{tabular}
\label{tab:confint1}
\end{table}
Given the estimated parameters, the derived ratios are $\hat{th}_{self} = 0.24$, $\hat{th}_{int} = 0.26$, $\hat{pa}_{int} = 0.25$, $\hat{pa}_{self} = 0.25$. The ratios indicate a highly balanced therapist-patient interaction in terms of both input and output energy. The therapist and patient are equally active in influencing the overall dynamics of the session, and equally receptive to the other's motions. This shows a strong mutual influence and engagement within the session, which can be indicative of a highly collaborative therapeutic process.

\begin{figure}[!ht]
\centering
\includegraphics[width=0.6\textwidth]{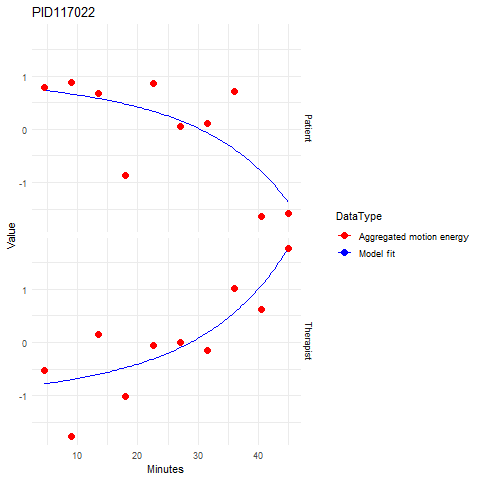}
hat\caption{\label{fig:fit2} Patient ID 117022: This figure depicts the aggregated motion energy data for both the therapist and patient, with a superimposed model fit. The model fit curve (shown in blue) demonstrates the performance of the model in capturing the patterns of the actual data, represented by red points. The plot provides a visual assessment of the model's accuracy in predicting motion energy trends.}
\end{figure}

Analyzing the second case study, Patient ID 117022, the parameter estimates were calculated as $\hat{\alpha} = 0.426$, $\hat{\beta} = 0.484$, $\hat{\gamma} = -0.313$, and $\hat{\delta} = -0.349$. The 95\% confidence intervals for the parameter estimates are displayed in Table \ref{tab:confint2}. Parameters $\alpha$ and $\beta$ have positive estimates, indicating a positive effect in the corresponding variables. The parameter $\delta$ is negative, reflecting a damping or regulatory effect. The small differences between the lower and upper bounds of the confidence intervals suggest a high degree of precision in the estimation of these parameters, reinforcing the reliability of our model. Based on the estimated values, the derived ratios are $\hat{th}_{self} = 0.27$, $\hat{th}_{int} = 0.30$,  $\hat{pa}_{int} = 0.20$, $\hat{pa}_{self} = 0.23$. 
Unlike the previous case study, here the eigenvalues are real and distinct, indicating a non-oscillatory, exponential pattern in the interaction dynamics, see Figure~\ref{fig:fit2}. 
\begin{table}[ht]
\centering
\caption{Patient ID 117022: 95\% Confidence Intervals for Parameter Estimates.}
\begin{tabular}{lcccc}
\hline
Parameter & Estimate & Lower Bound & Upper Bound \\
\hline
$\alpha$ & 0.426 & 0.423 & 0.428 \\
$\beta$ & 0.484 & 0.482 & 0.533 \\
$\gamma$ & -0.313 & -0.315 & -0.311 \\
$\delta$ & -0.349 & -0.352 & -0.347 \\
\hline
\end{tabular}
\label{tab:confint2}
\end{table}


Last, we analyze the session of Patient ID 117105. The model fitting procedure yielded parameter estimates of $\hat{\alpha}=-0.094$, $\hat{\beta}=-0.323$, $\hat{\gamma}=0.221$, and $\hat{\delta}=0.063$.
\begin{figure}[!ht]
\centering
\includegraphics[width=0.6\textwidth]{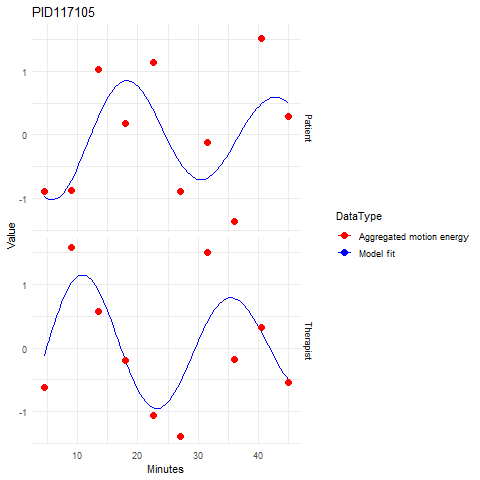}
\caption{\label{fig:fit3} Patient ID 117105: Time series plot illustrating the model fit to the aggregated motion energy data during a psychotherapy session. The red points represent the aggregated motion energy data for both therapist and patient, while the blue line corresponds to the model's fitted values. The plot reveals the model's ability to capture the dynamics of alternating periods of high and low activity. The leader-follower dynamic is also evident, with the modeled therapist's movements often preceding the patient's responses.}
\end{figure}

The 95\% confidence intervals for the parameter estimates are displayed in Table \ref{tab:confint3}. The parameters $\alpha$ and $\beta$ have negative estimates, while $\gamma$ and $\delta$ are positive. The confidence intervals here seem to be wider than in previous case studies. Furthermore, the confidence interval for $\delta$ includes zero, implying the parameter is not statistically significant at the 0.05 level. This might suggest that the damping effect of the patient is not as influential as other dynamics in this particular model.

\begin{table}[ht]
\centering
\caption{Patient ID 117105: 95\% Confidence Intervals for Parameter Estimates.}
\begin{tabular}{lcccc}
\hline
Parameter & Estimate & Lower Bound & Upper Bound \\
\hline
$\alpha$ & -0.094 & -0.203 & -0.007 \\
$\beta$ & -0.323 & -0.465 & -0.230 \\
$\gamma$ & 0.221 & 0.155 & 0.316 \\
$\delta$ & 0.063 & -0.024 & 0.173 \\
\hline
\end{tabular}
\label{tab:confint3}
\end{table}

The corresponding ratio values are $\hat {th}_{self}=0.14$, $\hat {th}_{int}=0.46$, $\hat {pa}_{int}=0.31$, and $\hat {pa}_{self}=0.09$. The eigenvalues in this case are complex conjugates, which suggests an oscillatory pattern in the dynamics of motion energy. Upon examining the time series for the therapist and patient in Figure~\ref{fig:fit3}, clear oscillations can be observed with alternating periods of high and low activity. A close inspection of the oscillatory pattern suggests that the therapist's movements often precede the patient's, indicative of a leader-follower dynamic. The patient's responses seem to be influenced by the therapist's actions, suggesting a reactive role. This observation aligns with the calculated interaction ratios which is higher for the therapist. A plausible interpretation is that the patient is responding more to the therapist's cues rather than initiating interactions.
 

It is important to emphasize that the above analysis and interpretations are based solely on motion energy data and should be considered with caution. The precise content or context of the therapy session was not taken into account, and further research is necessary to fully understand the complex interplay between these motion dynamics and other factors influencing the therapeutic process.

\section{Distinguishing between Trait and State Characteristics in Motion Dynamics}

In two recent papers \cite{zilcha2021toward}
and \cite{zilcha2022distinct} argue for the critical role of distinguishing between trait-like (stable) and state-like (dynamic) aspects of psychotherapy. They propose that this distinction provides a more personalized approach to psychotherapy, contributing to a deeper understanding of therapeutic change and the patient-therapist alliance. These works underline the necessity of considering both enduring attributes and momentary states when examining psychotherapeutic processes. They provide a significant motivation for the following analysis, which aims to explore trait-like and state-like characteristics within the dynamics of patient 
 and therapist motion energy. Specifically, we propose that the self-damping/reinforcing ratio parameters may represent trait-like characteristics of therapists and patients, while the interaction parameters may reflect state-like features of the therapeutic process. We analyzed the sessions of an experienced therapist spanning the years 2015-2018, finding compelling evidence for this proposed distinction. Notably, the therapist in question remained the same individual throughout this period, providing a consistent reference point for our analysis.

\subsection{Self-damping/reinforcing ratio}

\begin{figure}[!ht]
\centering
\includegraphics[width=0.6\textwidth]{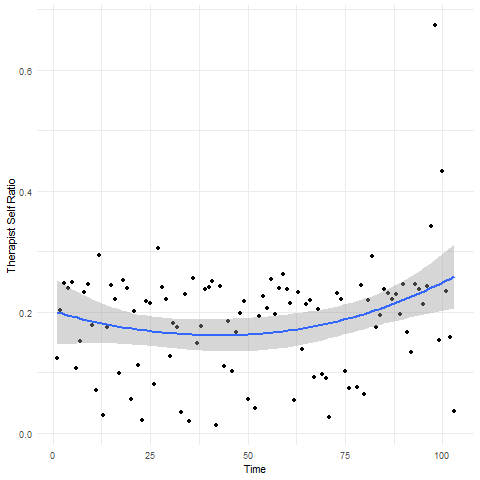}
\caption{\label{fig:th_self} Therapist's self-damping/reinforcing factor over time. The loess smoothing line illustrates the stable trend of the therapist's self factor over the period of 2015 to 2018.}
\end{figure}

\begin{figure}[!ht]
\centering
\includegraphics[width=0.6\textwidth]{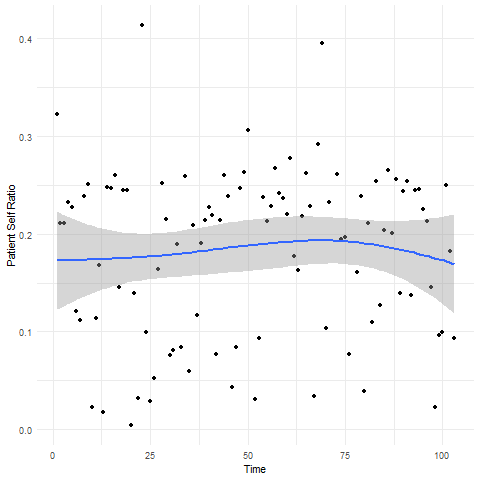}
\caption{\label{fig:pa_self} Patients' self-damping/reinforcing factor over the period of 2015 to 2018.}
\end{figure}

In Figure~\ref{fig:th_self} we can see that the therapist's self-damping/reinforcing parameter remained stable throughout this period, with a small positive trend towards the end of the period, maybe due to an outlier. Overall this parameter is around the value 0.2, namely 20\% of the session dynamics.   A linear regression model showed no significant trend, indicating that this parameter, and by implication the trait it represents, remains consistent over time. This suggests that certain aspects of a therapist's non-verbal communication style, such as his natural rhythm, may remain relatively fixed over time, functioning as a kind of therapeutic 'signature'. 
Further evidence for this distinction comes from the analysis of the patient's parameters over time as displayed in Figure~\ref{fig:pa_self}. It's important to note that unlike the therapist, the patients are different individuals, as the data is drawn solely from intake interviews. Despite this variability, the patient's self parameter showed no significant trend over time. The loess smoothing line oscillates around a value of 0.2, again about 20\% of the session dynamics, suggesting that this variability may be attributed to differences between patients, rather than a directional shift over time. This aligns with the notion that self-damping/reinforcing might represent a trait characteristic, and underscores the complexity of individual differences in psychotherapy dynamics.
\subsection{Interaction ratio}

\begin{figure}[!ht]
\centering
\includegraphics[width=0.6\textwidth]{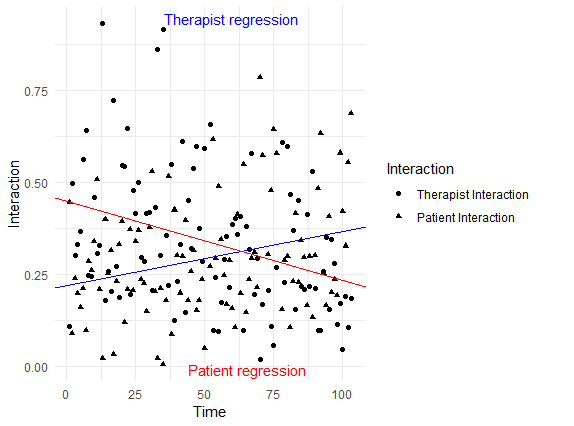}
\caption{\label{fig:inter_comp} Therapist and patient interaction over time. The regression lines illustrate the increasing trend in therapist interaction and the decreasing trend in patient interaction over the period of 2015 to 2018.}
\end{figure}

We now analyze the interaction ratio of both therapist and patient. The therapist's interaction parameter showed a significant positive trend (p-value = 0.005), while the patient's interaction parameter showed a significant negative trend (p-value = 0.0006). Figure~\ref{fig:inter_comp}  visualizes the interaction levels of the therapist and patients over time, with linear regression lines highlighting the underlying trends. The therapist's interaction shows a slight upward trend, while the patients' interaction reveals a downward trend. This inverse relationship between the therapist and patients' interaction levels could potentially suggest a shift in the dynamics of therapy sessions over time, with the therapist becoming more active and patients becoming less engaged in their interaction.
This may suggests an increased responsiveness to the patient's non-verbal cues over time, reflecting a possible evolution in therapeutic style or increased sensitivity to the patient's non-verbal expressions.
In both cases the model's R-squared value is about 7.8\%-11.5\%, suggesting that approximately 10\% of the variation in the interaction ratio
can be explained by the time of the session. This is not a very large amount of explained variance, suggesting that other factors not included in the model might also be influencing the interaction parameter over time. 

Notably, \cite{ramseyer2020motion} reported findings of decreasing synchrony between this specific experienced therapist and patients over time, suggesting that our model captures key features of therapeutic dynamics. Indeed, this intriguing parallel development of the interaction parameters - increasing for the therapist while decreasing for the patient - may jointly suggest a shift in the dyadic synchrony over time. This shift is not merely reflected in the  synchrony measure, but is rooted in the underlying mechanisms of interaction and self-damping/reinforcing that our model captures. These changes in the therapist and patient interactions not only reflect the individual engagement levels but also the interplay of these components, contributing to the overall synchrony in the dyad. This finding further substantiating that our mechanistic model captures critical aspects of the evolving therapeutic dynamics.

In summary, our results suggest that both trait-like and state-like characteristics play a role in  motion dynamics within psychotherapy sessions. The stability of the self-damping/reinforcing parameters highlights the enduring influence of individual traits, while the variability of the interaction parameters underscores the dynamic, evolving nature of the therapeutic process. Specifically, based on analyzing the specific data we have, it seems that the self-damping/reinforcing trait-like characteristics are 'responsible' to about 40\% of the motion dynamics, while the interaction state-like characteristics for about 60\%. These insights further our understanding of psychotherapy dynamics and underscore the potential of our modeling approach for capturing the complexity of therapeutic interactions. Future research should continue to explore these dynamics, investigating the potential impacts of these trait-like and state-like characteristics on therapy outcomes and the professional development of therapists.

\section{Discussion and Conclusions}

This study introduces a pioneering mathematical and statistical framework for exploring the dynamics of psychotherapy, with an emphasis on the analysis of motion energy data obtained during therapy sessions. Our methodology, anchored in a system of coupled linear ordinary differential equations, delves into the intricate mechanisms propelling motion dynamics in therapeutic dyads. Furthermore, the ability of our approach to manage measurement errors and deliver trustworthy parameter estimates and confidence intervals highlights its accuracy and reliability. By providing a more comprehensive understanding of therapeutic dynamics, this research opens the door to advanced data-driven insights in the field of psychotherapy.

Through the analysis of three case studies, we demonstrated the practical utility of our model. By transforming raw motion energy data into interpretable narratives of non-verbal communication patterns, we identified meaningful dynamics and roles within therapist-patient dyads. This ability to extract actionable insights from motion energy data showcases its potential in revealing unique perspectives on psychotherapy dynamics. Our in-depth investigation also brought forth the importance of distinguishing between trait and state characteristics of the dyadic interaction dynamics, taking inspiration from recent works in the field of psychotherapy research. We observed how the trait-like and state-like characteristics of therapist and patient interactions manifested in the therapy sessions, providing a nuanced understanding of the dynamic interplay between consistent patterns and moment-to-moment fluctuations in non-verbal communication. The insights gained from this analysis not only shed light on the nuances of therapist-patient dynamics but also underscore the potential for a broader applicability of our mechanistic model to other dyadic interactions. 

While the mechanistic modeling approach provides significant insights into the dynamics of psychotherapy sessions, it is important to underscore the fundamental difference between mechanism and causality. By examining the patterns of interaction ratios over time for both the experienced therapist and patients, we observe noticeable changes, indicating evolving dynamics. However, attributing causality based solely on these changes could lead to multiple, and potentially conflicting interpretations. For instance, one possible interpretation could be that as the therapist gains more experience, he tends to engage with more challenging patients, necessitating a higher degree of involvement on his part. Another plausible explanation could be that as the therapist's experience grows, he becomes more proactive in the therapeutic interaction, potentially overshadowing the patient's participation and thus reducing the degree of synchrony.

Both explanations are plausible based on the available motion energy data. Nevertheless, they represent contrasting views on the cause-and-effect dynamics at play: the first suggests a reaction to the changing patient population, while the second implies an inherent change in the therapist's approach over time. The key takeaway is that the observed mechanism – the changing dynamics of interaction ratios over time – does not inherently reveal the underlying causality. Further research, perhaps incorporating additional data sources or methods, would be needed to untangle the complex web of cause and effect in these interactions. Mechanistic modeling thus serves as a powerful tool for revealing patterns and generating hypotheses, but it must be complemented with careful interpretation and further investigatory work to draw robust causal inferences. Indeed, our motion-based analyses, while offering a new perspective on psychotherapy dynamics, do not provide direct insights into therapy session content or participant subjective experiences. Traditional data sources, such as session transcripts or self-report measures, are better equipped to capture these elements.

Future research directions offer exciting prospects. Incorporating process noise within our models could yield a more comprehensive understanding of psychotherapy dynamics. Applying our methodology to various types of dyadic interactions or different therapeutic modalities is another promising direction. Furthermore, delving deeper into the relationship between motion dynamics and therapy outcomes could enhance our understanding of different therapeutic interventions' effectiveness.

In conclusion, our research marks a significant step forward in the quantitative analysis of non-verbal synchrony in psychotherapy. By adopting a mechanistic understanding of these dynamics, we pave the way for further advancements in this field. The potential to derive meaningful insights from data collected non-intrusively and analyzed objectively opens up new avenues in therapeutic research and practice.
\bibliography{main}

\end{document}